\documentclass[conference]{IEEEtran}
\IEEEoverridecommandlockouts
\usepackage{cite}
\usepackage{amsmath,amssymb,amsfonts}
\usepackage{tcolorbox}

\usepackage{array}
\usepackage{tabularx}
\usepackage{algorithmic}
\usepackage{graphicx}
\usepackage{textcomp}
\usepackage{csquotes}

\usepackage{xcolor}
\def\BibTeX{{\rm B\kern-.05em{\sc i\kern-.025em b}\kern-.08em
    T\kern-.1667em\lower.7ex\hbox{E}\kern-.125emX}}

\makeatletter
\newcommand{\linebreakand}{%
\end{@IEEEauthorhalign}
\hfill\mbox{}\par
\mbox{}\hfill\begin{@IEEEauthorhalign}
}
\makeatother

\begin{document}

\title{Explainability as a Compliance Requirement: What Regulated Industries Need from AI Tools for Design Artifact Generation}

\author{
    \IEEEauthorblockN{ Syed Tauhid Ullah Shah}
    \IEEEauthorblockA{\textit{Schulich School of Engineering} \\
    \textit{University of Calgary}\\
    Calgary, Canada \\
    syed.tauhiidullahshah@ucalgary.ca}
    \and
    \IEEEauthorblockN{Mohammad Hussein}
    \IEEEauthorblockA{\textit{Schulich School of Engineering} \\
    \textit{University of Calgary}\\
    Calgary, Canada \\
    mohammad.hussein@ucalgary.ca }
    \and
    \IEEEauthorblockN{ Ann Barcomb\textsuperscript{*}}
    \IEEEauthorblockA{\textit{Schulich School of Engineering} \\
    \textit{University of Calgary}\\
    Calgary, Canada \\
    ann.barcomb@ucalgary.ca }
    \linebreakand
    \IEEEauthorblockN{Mohammad Moshirpour}
    \IEEEauthorblockA{\textit{Department of Informatics} \\
    %\textit{Donald Bren School of Information and Computer Sciences} \\
    \textit{University of California at Irvine}\\
    Irvine, California, USA \\
    mmoshirpour@uci.edu }
}

\maketitle

\begin{abstract}
Artificial Intelligence (AI) tools for automating design artifact generation are increasingly used in Requirements Engineering (RE) to transform textual requirements into structured diagrams and models. While these AI tools, particularly those based on Natural Language Processing (NLP), promise to improve efficiency, their adoption remains limited in regulated industries where transparency and traceability are essential. In this paper, we investigate the explainability gap in AI-driven design artifact generation through semi-structured interviews with ten practitioners from safety-critical industries. We examine how current AI-based tools are integrated into workflows and the challenges arising from their lack of explainability. We also explore mitigation strategies, their impact on project outcomes, and features needed to improve usability. Our findings reveal that non-explainable AI outputs necessitate extensive manual validation, reduce stakeholder trust, struggle to handle domain-specific terminology, disrupt team collaboration, and introduce regulatory compliance risks, often negating the anticipated efficiency benefits. To address these issues, we identify key improvements, including source tracing, providing clear justifications for tool-generated decisions, supporting domain-specific adaptation, and enabling compliance validation.  This study outlines a practical roadmap for improving the transparency, reliability, and applicability of AI tools in requirements engineering workflows, particularly in regulated and safety-critical environments where explainability is crucial for adoption and certification.
\end{abstract}


\section{Introduction}

Requirements Engineering (RE) involves defining and managing processes to develop systems that fulfill user and stakeholder expectations \cite{maalej2016automated}. It encompasses tasks such as eliciting, analyzing, documenting, and validating requirements to ensure that systems meet functional and regulatory goals \cite{nuseibeh2000requirements}. As systems grow more complex and regulated, RE plays a critical role in maintaining traceability, ensuring compliance, and supporting safety certification efforts \cite{wang2016requirements}. These demands make RE not only a foundation for system correctness but also a regulatory and organizational imperative in safety-critical domains \cite{wagner2019requirements}. One of the main challenges in RE is converting unstructured textual requirements into structured design artifacts \cite{ghazi2017challenges}. This model transformation process produces outputs like class diagrams, sequence diagrams, activity diagrams, and deployment diagrams \cite{pastor2011requirements}. Creating these artifacts manually is time-consuming and error-prone, especially in large projects with multiple teams~\cite{aurum2005engineering}. To reduce this effort, AI tools are used in RE workflows to automate the design artifact generation task~\cite{ferrari2019ambiguity, wang2024software}. These tools improve consistency and reduce manual effort, allowing engineers to focus more on system-level analysis and decision-making~\cite{zhao2021natural}.

Despite their potential to improve RE efficiency, AI tools have seen limited adoption, mainly due to concerns about explainability and resulting trust issues~\cite{zhao2021natural}. Tool outputs are often opaque and require extensive manual validation to ensure correctness, which can offset the intended efficiency gains~\cite{ezzini2023aiqa}. This lack of explainability undermines trust and limits the broader application of AI tools in RE~\cite{zamani2021ml}. Addressing these explainability challenges is essential to unlocking the full potential of AI tools in RE~\cite{hou2024llms}. Although AI tools are increasingly used in software development, there remains a research gap on how explainability affects their use in RE, particularly in regulated industries where transparency is required~\cite{hou2024llms}. Existing studies focus on the technical capabilities of AI tools for requirements analysis, while the practical challenges of integrating them into real-world RE workflows, where trust, explainability, and regulatory compliance are essential, remain underexplored.

To address this gap, we adopt the formal definition of explainability provided by Chazette et al.~\cite{chazette2021exploring}: ``A system S is explainable with respect to an aspect X of S relative to an addressee A in context C if and only if there is an entity E (the explainer) who, by giving a corpus of information I (the explanation of X), enables A to understand X of S in C.'' In the context of AI tools for RE, this means that for a tool to be considered explainable, it must provide sufficient information (I) that enables requirements engineers (A) to understand why specific design artifacts were generated (X) within their particular project and regulatory context (C). Furthermore, as Chazette and Schneider~\cite{chazette2020explainability} establish, explainability functions as a non-functional requirement that serves as ``a means to achieve transparency,'' distinguishing it from related concepts like interpretability, which focuses more on understanding the technical decision-making process of the model itself.

%\ab{One of the key concerns of reviewers was the lack of a definition for explainability. You do have one in your survey. For a research paper, I strongly suggest that you do not attempt to paraphrase to create your own definition, but instead directly quote someone else. A lot of work goes into creating a high quality definition, therefore it is best to directly quote rather than attempt your own, when dealing with a critical concept which is core to the paper.} \textcolor{blue}{added the paragraph above}

In this study, we investigated the explainability gap in AI tools used in RE, focusing on their role in preliminary design integration within regulated industries. We conducted semi-structured interviews with ten RE practitioners from sectors including aerospace, automotive, medical devices, energy, telecommunications, and industrial automation. Based on these interviews, we identified key challenges related to non-explainable tool outputs, such as validation difficulties, trust issues, domain knowledge gaps, workflow disruptions, and compliance risks. We also documented the mitigation strategies practitioners adopted to work around these issues and analyzed the hidden costs and inefficiencies they introduced. Additionally, we examined the broader organizational impacts, including effects on collaboration, stakeholder communication, project delays, and regulatory processes. Finally, we synthesized these findings into a set of actionable recommendations for improving the explainability and usability of AI tools in RE, offering guidance for both tool developers and organizations operating in safety-critical environments.

Our work is guided by the following Research Questions (RQs), addressed through a qualitative study involving interviews with practitioners across multiple regulated industries. This research is motivated by a core paradox: while AI tools offer substantial efficiency gains for RE, their adoption remains limited in precisely those sectors where they are most needed, domains governed by strict regulatory and safety standards. Unlike prior work that focuses primarily on technical capabilities or proposes new algorithms, our study offers the first systematic investigation of how explainability affects real-world adoption in such contexts. We move beyond lab-based evaluations to examine practitioners’ lived experiences, uncovering not only technical challenges but also organizational workarounds, hidden compliance costs, and socio-technical frictions often overlooked in existing literature. This practitioner-centered lens reveals why technically capable tools fail in practice and identifies the specific explainability features required to enable their successful adoption in safety-critical RE settings.% \ab{The introduction could do a better job of motivation and novelty. Be more explicit. The motivation is given somewhat in paragraph 2 where you talk about limited adoption. You can add to this paragraph by giving your explicit motivation for the research here. The last sentence of that paragraph touches on novelty, but it could be more explicit, as could the following paragraph (paragraph 3).} \textcolor{blue}{Extended this paragraph. Also, the length of the paper should be 10 pages with references as well. So, after removing the comments, it is still 11 pages. Thus, I would kindly request to kindly point out what I need to cut}

\begin{itemize}

    \item \textbf{RQ1}: \textit{Which AI tools are currently used for preliminary design artifacts in RE across different industries?}
To establish a baseline, this question identifies which AI tools practitioners use for generating design artifacts, ranging from commercial platforms to in-house solutions, providing context for their capabilities and limitations across industrial settings. %\ab{The RQ doesn't match the explanation. Is the question which tools or how are those tools used?} \textcolor{blue}{fixed}

    \item \textbf{RQ2}: \textit{What challenges do RE practitioners face due to the lack of explainability in AI tools?}
    A major barrier to adoption is the opaqueness of AI-generated outputs. This question investigates how the lack of explainability affects the ability of practitioners to trust, validate, and refine tool-generated artifacts, as well as its broader impact on efficiency and decision-making.

    \item \textbf{RQ3}: \textit{How do organizations adapt to the explainability limitations of AI tools in RE?}
    When AI tools fail to provide clear justifications for their outputs, practitioners develop workarounds to ensure reliability. This inquiry focuses on these strategies, emphasizing the workload involved in managing non-explainable outputs and the challenges that disrupt tool adoption.

    \item \textbf{RQ4}: \textit{What are the broader impacts of non-explainable AI tools on RE workflows, team collaboration, and project success?}
    This question investigates the broader consequences of non-explainable AI tools in RE. It analyzes how these tools affect workflow effectiveness, team coordination, and stakeholder communication. It also examines their impact on compliance with industry regulations. Further, it considers critical project outcomes such as cost, time, and risk management.
    
    \item \textbf{RQ5}: \textit{What key features would improve the explainability and usability of AI tools in RE?}
    Building on earlier insights, this RQ focuses on solutions, highlighting practitioner-driven requirements to improve tool transparency, traceability, and reliability to better support RE workflows and regulatory needs.

\end{itemize}

\section{Related Work}

Transforming natural language requirements into design artifacts is a core challenge in RE. Machine Learning (ML) methods have advanced significantly, particularly in domain concept extraction and UML diagram generation \cite{camara2023assessment, saini2022automated, jahan2021generating, arora2019active}, and recent Large Language Model (LLM) studies have shown potential in goal-modeling and UML generation \cite{norheim2024challenges}.  However, systematic evaluations reveal persistent limitations: these tools struggle with completeness, consistency, and semantic accuracy when processing complex natural language \cite{ferrari2018detecting}.

Despite technical progress, there remains a substantial gap between research prototypes and industrial adoption. Reviews indicate that 67–87\% of AI tools for RE are confined to lab settings, with merely 7–13\% tested in practical contexts \cite{zhao2021natural, necula2024systematic}. Explainability issues emerge as key barriers, with practitioners valuing transparency as highly as accuracy \cite{jiarpakdee2021practitioners}. Although theoretical frameworks exist for transparency requirements and trace link justifications~\cite{balasubramaniam2023transparency,habiba2022can}, their validation remains limited to controlled environments rather than real-world contexts where opacity disrupts validation workflows and stakeholder trust~\cite{jiarpakdee2021explainable}. Broader adoption studies identify recurring issues of trust, customization, and integration challenges \cite{li2024ai, russo2024navigating}, yet overlook how explainability affects artifact generation, potentially impacting safety.

Limited industrial case studies provide valuable but partial insights. For instance, explainable classifiers in railway and automotive sectors have shown efficiency and compliance improvements \cite{fazelnia2024railway}. However, even successful deployments entail extensive manual validation efforts, suggesting hidden inefficiencies \cite{elahidoost2024visma}. These studies do not explore variations in explainability needs across industries or examine practitioner strategies and associated costs.

While these industrial studies document technical implementations and outcomes, they overlook a crucial dimension: the practitioner perspective on managing explainability gaps in their daily workflows. Specialized reviews document interpretability limitations across various RE applications including software testing and user stories~\cite{raharjana2021user}, and traceability~\cite{pauzi2023applications}. These reviews consistently report shallow practitioner understanding without investigating the underlying causes~\cite{sonbol2022use}. Yet no prior work systematically examines how practitioners detect and correct errors when tools provide no justification for their outputs, or investigates the actual strategies teams employ to maintain regulatory compliance when tool reasoning remains opaque. Furthermore, existing studies focus exclusively on active users, missing critical insights from practitioners who have abandoned these tools due to explainability concerns. 

Our study presents the first cross-industry investigation into explainability challenges in real-world design artifact generation. We go beyond technical limitations to examine how organizations adapt in the absence of explainability, including the hidden costs, manual workarounds, and collaboration issues that arise. These factors ultimately determine whether AI tools provide meaningful value in RE workflows or become costly obstacles to their adoption and effectiveness.

\section{Methodology}

\subsection{Population and Sampling Strategy}

We interviewed ten software requirements engineers from highly regulated industries. 
Participants were selected via convenience sampling \cite{emerson2015convenience} through LinkedIn outreach, with a minimum requirement of two years of RE experience. Participants were anonymized in accordance with the study's approved ethics, and are referred to by alias (A-J) throughout the paper.
Most participants (A–G) actively used AI tools, while others (H–J) had discontinued use, allowing us to capture both adoption and abandonment perspectives. Table~\ref{table:participant_profiles} summarizes their profiles.

\begingroup
\setlength{\abovedisplayskip}{0pt}
\setlength{\belowdisplayskip}{0pt}
\setlength{\textfloatsep}{0pt}
\begin{table}[h]
\caption{Participant Profiles}
\label{table:participant_profiles}
\centering
\resizebox{\columnwidth}{!}{%
\begin{tabular}{|l|r|l|}
\hline
\textbf{Participant} & \textbf{Years of Experience} & \textbf{Domain Expertise} \\ \hline
A & 20+ years & Aerospace Systems \\ \hline
B & 8 years & Autonomous Vehicle Systems \\ \hline
C & 10 years & Automotive Manufacturing \\ \hline
D & 8 years & Industrial Automation \\ \hline
E & 3 years & Medical Device Industry \\ \hline
F & 2 years & Energy and Utilities \\ \hline
G & 3 years & Telecom Infrastructure \\ \hline
H* & 5 years & Embedded Systems and Automation \\ \hline
I* & 6 years & Automotive Embedded Systems \\ \hline
J* & 5 years & Automotive Embedded Systems \\ \hline

\multicolumn{3}{l}{* Participants who previously used AI tools but discontinued their use.}
\end{tabular}%
}
\end{table}
\endgroup
\subsection{Data Collection}
We designed a four-part interview protocol aligned with our RQs. First, we asked about participants’ backgrounds, experience levels, and RE-related challenges. Next, we explored their use of AI tools, including tool types, workflow integration, and practical experiences. We then examined challenges arising from non-explainable outputs, focusing on validation and trust issues. Finally, we discussed desired improvements to enhance explainability and usability in AI tools. Each interview lasted 90 to 120 minutes and was conducted online to accommodate participants across locations. To elicit detailed insights, we encouraged participants to share specific examples from their professional experience. The interview script and codebook are available for replication\footnote{https://figshare.com/s/53ecc93379138814cf46}.

\subsection{Data Analysis}

\begin{figure}[h]
    \centering
    \includegraphics[width=\columnwidth]{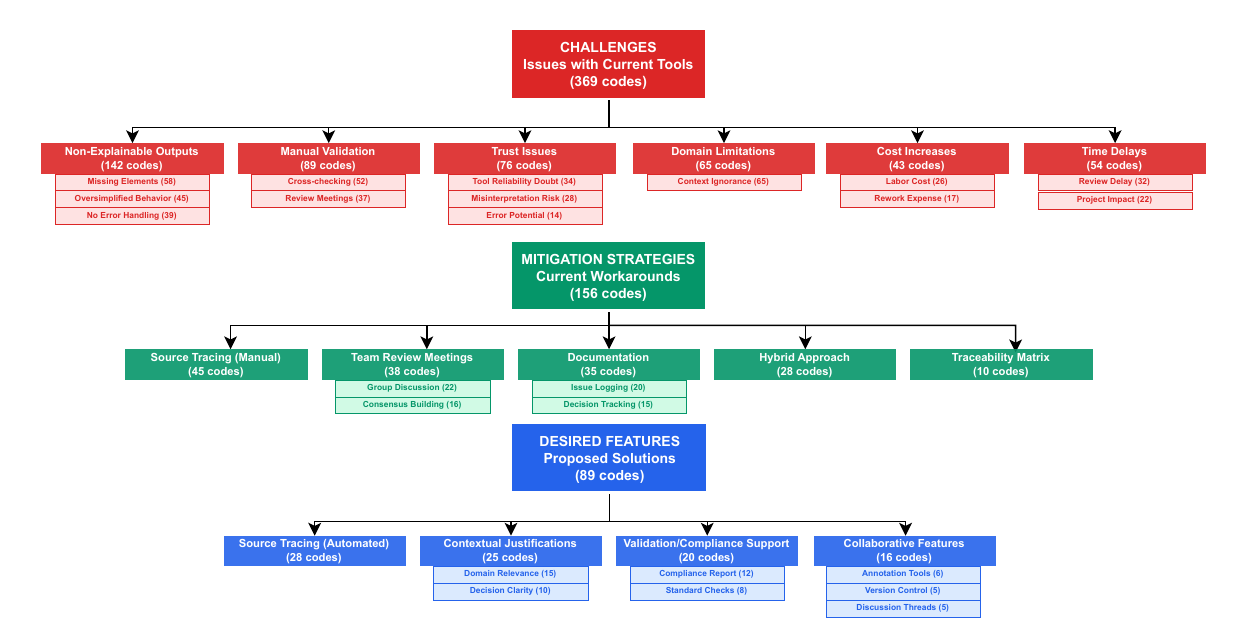}
    \caption{Thematic Map of Explainability Challenges in AI Tools for Requirements Engineering.}
    \label{fig:thematic_map}
\end{figure}

We used qualitative data analysis (QDA) with inductive coding principles from Charmaz’s grounded theory approach \cite{charmaz2006constructing} to identify key themes from the interviews. Two researchers independently coded transcripts in MaxQDA\footnote{https://www.maxqda.com/}, starting with codes aligned to our RQs and refining them as new insights emerged. We followed a hybrid approach where high-level categories were defined a priori, while low-level codes were developed inductively during analysis. To ensure consistency, we created a shared codebook and resolved disagreements through discussion. To assess the reliability of our coding, we calculated inter-coder agreement using Cohen’s Kappa, which was approximately 0.76, indicating a high level of consistency between the two coders.  After six interviews, no additional themes emerged, suggesting data saturation. In total, we identified 59 unique codes across 1163 coded excerpts. Figure~\ref{fig:thematic_map} presents the complete thematic structure derived from our qualitative analysis, illustrating the hierarchical relationship between main themes and subcodes with their respective frequencies. 

\section{Results}
\label{sec:results}
% \begingroup
% \setlength{\abovecaptionskip}{0pt} % Reduce space above caption
% \setlength{\belowcaptionskip}{0pt} % Reduce space below caption
% \setlength{\textfloatsep}{0pt}
% \begin{figure}[h]
%     \centering
%     \includegraphics[width=\columnwidth]{RE.pdf}
%     \caption{Conceptual Framework of Explainability in AI Tools for Requirements Engineering.}
%     \label{fig:explainability_flow}
% \end{figure}
% \endgroup

\subsection{Current Use of AI Tools for Preliminary Design Integration in RE (RQ1)}

Our RQ1 analysis draws on the \textit{Tool Usage}, \textit{Design Artifact Generation}, and \textit{Efficiency Gains} codes from our codebook. Interviews revealed that AI tools are integrated into RE workflows in different ways to support early design activities. As shown in Table~\ref{table:tool_usage}, participants used both commercial and custom-built tools to transform textual requirements into design diagrams. The most common outputs were UML class diagrams, sequence diagrams, state charts, and system architecture diagrams. Only one participant reported using a custom-built tool, reflecting the rarity of in-house solutions.

\begin{table}[h]
\caption{AI Tools and Design Artifacts Used by Participants}
\label{table:tool_usage}
\centering
\renewcommand{\arraystretch}{1.05}
\setlength{\tabcolsep}{3pt}
\scriptsize % Reduce font size to fit column width
\begin{tabular}{|p{2.7cm}|p{3.6cm}|p{1.6cm}|}
\hline
\textbf{Tool} & \textbf{Design Artifacts} & \textbf{Participants} \\
\hline
DiagramGPT & Class, Sequence, State, System Architecture, Control Flow Diagrams & E, F, G \\
\hline
MyMap.ai & Control Flow, State Charts, UML Diagrams & B, G, J \\
\hline
Eraser.io & Sequence, State Diagrams & A, F \\
\hline
Diagramly.ai & Control Flow, State Charts, UML Class and Sequence Diagrams & B, C \\
\hline
Miro AI & UML Class Diagrams, Sequence Flows & H \\
\hline
ChatUML & UML Class, Sequence Diagrams & I \\
\hline
ChatGPT & UML Class, Sequence Diagrams & C \\
\hline
PlantUML & UML Class, Sequence Diagrams & C \\
\hline
LLM-based internal tool & Interaction Sequence Diagrams & D \\
\hline
Internally developed AI & System Architecture, UML Diagrams, Sequence Flows & E \\
\hline
\end{tabular}
\end{table}

Participants reported two main reasons for using AI tools: saving time and managing large volumes of requirements.
Many participants found that AI tools helped speed up the early design phase by automating the creation of initial diagrams. Instead of manually constructing each design artifact, teams could generate draft versions within minutes. This allowed them to focus on refining and verifying outputs instead of spending time on basic diagram creation. Tasks that once took days or weeks were completed much faster. Some participants used AI tools to handle the increasing volume of requirements more efficiently. In projects with extensive documentation, AI-assisted workflows reduced the effort required for manual reviews, making it easier to maintain accuracy while managing complex systems. 

\subsection{Challenges Arising from Non-Explainability \textbf{RQ2} }

\subsubsection{Non-Explainable Outputs}
A common challenge with AI-generated design artifacts is their lack of transparency, reported by 9 out of 10 participants. They found outputs structured but lacking depth, leading to three key issues: missing system elements, oversimplified behavior representations, and no consideration for error handling.

A frequent concern was the omission of key system components without any explanation. Participant E noted, \textit{``The requirement clearly stated that the system should pause, but the tool didn’t explain why it left it out.''} Participants also noted that AI tools struggled to handle complex system interactions and dependencies. Rather than reasoning about interactions, they extracted text patterns and rearranged them into structured formats. This often led to diagrams that appeared correct but failed to capture dependencies, constraints, or real-world scenarios. As Participant H put it, \textit{``These tools don’t design systems, they rearrange words into structured formats. They miss fail-safes, ignore timing constraints, and don’t consider real-world interactions.''}

AI tools also failed to account for complex system behaviors. Participant I stated, \textit{The AI had zero understanding of system-level design. It wasn’t thinking like an architect, it was just drawing boxes and arrows based on whatever words it recognized in the requirements.''} The lack of transparency also affected collaboration. Engineers had to rely on trial and error to verify tool outputs. Participant C explained, \textit{``When we reviewed the sequence diagrams, we noticed missing interactions between the braking system and the central controller. The tool didn’t flag this omission or indicate why it had overlooked a key requirement.''} Explainable results have the potential to improve the usefulness: \textit{``Explainability is the biggest barrier to adoption''} (Participant B).
While these tools provide a starting point, their lack of explainability limits trust and usability, requiring extensive manual validation to avoid critical design flaws.

\subsubsection{Manual Validation}
Our analysis, based on the \textit{Manual Validation} code from our codebook, highlights the additional effort required to verify AI-generated design artifacts, reducing expected efficiency gains.  Participants across all domains reported that extensive manual validation was required to verify AI-generated design artifacts. This additional effort reduced the expected efficiency gains of these tools. Since the tools did not explain their reasoning, teams could not fully trust their outputs without careful review.  

To address this, teams developed validation workflows. Participants noted cross-checking tool-generated designs against the original requirements to ensure accuracy. When outputs appeared incorrect or unclear, engineers manually traced issues back to the source text. In some cases, formal review meetings were held, where flagged inconsistencies were analyzed with subject matter experts. Manual validation was particularly important in safety-critical environments, where missing or incorrect elements posed compliance risks. Participant C emphasized, \textit{``Every design decision must meet strict safety standards, so we manually check each diagram to ensure it complies with regulations like ISO 26262.''}  

These validation efforts had a major impact on workflows. Some participants reported spending weeks manually reviewing flagged interactions, leading to project delays. One participant F noted, \textit{``Situations like this slow us down. Instead of trusting the tool’s output, we have to treat it as a rough draft and spend time revalidating everything.''}  This showed a shift in understanding that while AI tools provide a starting point, their outputs require thorough validation.

\subsubsection{Trust Issues}

Our analysis is based on the \textit{Trust Issues} code from our codebook with subcodes (\textit{Tool Reliability Doubt}, \textit{Misinterpretation Risk}, and \textit{Error Potential}). Our respondents found it difficult to trust the outputs of AI tools and believed they required validation. As Participant C explained, \textit{``Over time, the team has become more skeptical. Instead of trusting the tool’s outputs, we see them as a rough starting point.''} In safety-critical industries, mistakes can have severe consequences. Participant C further explained, \textit{``If a tool makes a mistake, it could create a design flaw that affects safety compliance.''} To convince stakeholders that the design adheres to safety requirements, requirements engineers need to be able to explain the process:  
\textit{``We don't even show the AI-generated outputs to external stakeholders because we can't explain them properly''} (Participant F).

This lack of trust has created uncertainty around the use of AI tools. Participant A described the difficulty, \textit{``Right now, we're stuck in a cycle of trusting the tool, then verifying its output, then deciding whether to trust it again.''} The constant need for validation removes the primary benefit of automation, as Participant D observed, \textit{``honestly, it defeats the purpose of automation.''} This problem led some participants to give up using automation tools. Participant H said: \textit{``What should have been a tool to assist engineers became an extra step in the process, one that didn’t add any value.''} 

\subsubsection{Domain Limitations}
Our analysis for this phase was based on the \textit{Domain Limitations} code from our codebook with the subcode \textit{Context Ignorance}, which showed that AI tools struggle with industry-specific aspects of RE, affecting their usefulness across different fields. One key issue was the tools’ inability to understand domain-specific architectures and system behaviors. In automotive systems, for example, tools generated designs that misrepresented how components interact. As Participant I explained, \textit{``The AI assumed the ESC directly controlled the braking system, but in reality, that’s not how modern vehicle architectures work.''} 

Another challenge was domain-specific terminology. Tools often misinterpret industry-specific terms by incorrectly applying general software definitions. Participant F provided an example: \textit{``In a power grid, a Load Balancer redistributes power loads dynamically. But in general software development, a load balancer refers to network traffic distribution, which is completely different.''} These misinterpretations led to incorrect classifications and relationships in system diagrams.

AI tools also failed to incorporate industry standards and regulatory requirements. Participant D noted, \textit{``Our tool wasn’t designed for industrial automation. It lacks understanding of our best practices, compliance needs, and industry standards.''} This created compliance risks, particularly in safety-critical industries where regulatory adherence is essential.

Participants attributed these limitations to the tools' keyword-based approach to processing requirements rather than a deeper understanding of system functions. \textit{``These tools assume that just because they can extract keywords from a requirement, they understand the system. But they don’t.''}, (Participant H). A major reason for these issues appeared to be the broad training data used for commercial AI tools. As Participant F observed, \textit{``Right now, the biggest issue is that the tools are too generic. They’re trained on broad datasets, but they don’t have deep knowledge of specific industries.''} This gap between tool capabilities and domain-specific needs made adoption difficult in fields requiring failure mode analysis and redundancy mechanisms.

\subsubsection{Time Delays}
These are downstream effects of non-explainability rather than standalone challenges. Our analysis, based on the \textit{Time Delays} code from our codebook with subcodes (\textit{Review Delay} and \textit{Project Impact}) found that the lack of explainability in AI tools often led to time delays, reducing the efficiency benefits these tools were meant to provide. These delays occurred at different stages of the RE workflow.

A major cause of delays was the need for extensive manual validation. Since the tools did not explain how they arrived at their results, engineering teams had to introduce extra validation steps, which extended project timelines. As Participant G explained, \textit{``In the network slicing project, our initial timeline was estimated for two weeks for the design phase. Instead, it took nearly six weeks because of all the manual validation required.''} Some organizations even added formal \textit{``Tool Output Validation''} phases to their project plans to account for this extra work.

The issue becomes more difficult when tools produce incorrect or incomplete outputs. In these cases, teams not only had to identify errors but also determine why the tool made those mistakes. Participant E described the challenge: \textit{``We had to manually go back and reconstruct the entire workflow to see what logic it might have used.''} This additional effort was necessary to prevent similar errors in future workflows. Some teams spent as much or more time verifying tool outputs as they would have spent creating designs manually. Participant D expressed this as: \textit{``We thought the tool would speed things up, but we spend more time double-checking its outputs than we would doing the work manually from the start.''}

These delays also affected later stages of development, particularly in industries with strict regulatory requirements. For teams working under tight deadlines, this created additional project management challenges. Participants emphasized that these delays were not caused by AI tools themselves but rather by their lack of explainability.

\subsubsection{Cost Increases}
Our analysis, based on the \textit{Cost Increases} code from our codebook with subcodes (\textit{Labor Cost} and \textit{Rework Expense}) found that the lack of explainability in AI tools led to higher costs, particularly due to labor and resource allocation. The biggest financial impact came from the extra labor required for manual validation. As Participant C explained, \textit{``The biggest cost factor is engineering time. Every time we have to manually verify a tool's output, it adds extra hours to the workload.''} 

Rework costs were another concern. When tools produced incomplete or incorrect outputs, teams had to spend additional time fixing them.  In some cases, teams had to rebuild design artifacts entirely. Participant F described, \textit{``We had to manually verify every output, compare it against the requirements, and rework the diagram from scratch.''} 
Costs increased further when errors were not caught early. If mistakes made it into later development stages, fixing them became much more expensive. Participant C highlighted, \textit{``If we catch the mistake early during requirements analysis, we might lose a few days correcting the model. But if the issue makes it to implementation or testing, it can require weeks of rework, which significantly drives up costs.''} 

In regulated industries, these costs were compounded by compliance risks. Participant E explained, \textit{``If we hadn't manually reviewed the output, that flaw could have made its way into the final design, which would be a compliance and safety risk.''} Failing to detect such issues early could result in certification delays, regulatory fines, or even product recalls, and financial risks beyond direct engineering costs. 

These challenges create inefficiencies that disrupt workflows, delay projects, and pose compliance risks. The next subsection~\ref{subsec:rq3} explores mitigation strategies in detail.
\subsection{\textbf{RQ3}: \label{subsec:rq3} Strategies to Mitigate Challenges}
To address the explainability challenges of AI tools in RE, practitioners have implemented various workarounds to enhance traceability, validation, and usability. This section outlines key strategies identified through participant interviews.

\subsubsection{Source Tracing (Manual)}

Our analysis, based on the \textit{Source Tracing (Manual)} code from our codebook with 30 coded references across participants, found that practitioners manually traced outputs back to their original requirements. This process involved cross-checking tool-generated diagrams against source documents to verify accuracy. Participants reported comparison methods to assist with validation. As Participant C explained, \textit{``We placed the tool-generated diagram side by side with a manually created reference diagram based on our existing system architecture.''} When discrepancies arose, engineers manually traced the reasoning of the tool. This process often required extensive back-and-forth reviews. Participant B described, \textit{``We compared diagrams with textual requirements, double-checked constraints, and held meetings to determine if the issue was a tool error or unclear wording.''}

The time-intensive nature of manual tracing reduced the efficiency benefits of AI tools. Participant D highlighted the impact, stating, \textit{``If the tool could provide direct references to industry standards, we wouldn’t have to waste time hunting down the reasoning ourselves. That alone would save us weeks of effort.''} Directly linking outputs to inputs would reduce validation time, improve trust, and address a major explainability gap in current AI tools for RE.

\subsubsection{Team Review Meetings}

Our analysis is based on the \textit{Team Review Meetings} code from our codebook with subcodes (\textit{Group Discussion} and \textit{Consensus Building}), highlighting how organizations addressed the lack of explainability in AI tools by holding team review meetings to validate tool-generated outputs. As Participant B explained, \textit{``We hold review meetings where engineers, designers, and safety experts go through the tool's output together.''} During these meetings, teams compared tool outputs with original requirements, often using visual side-by-side comparisons to identify discrepancies. These sessions also helped teams align on design decisions, compensating for the tool’s lack of explainability by using human expertise. However, participants noted that multiple review sessions were often required, leading to project delays. 

\subsubsection{Documentation}

Based on the \textit{Documentation} code from our codebook with subcodes (\textit{Issue Logging} and \textit{Decision Tracking}), organizations developed documentation practices to track validation processes and decision-making, addressing the explainability limitations of AI tools. Teams categorized common errors to identify patterns in tool behavior. As Participant C explained, \textit{``We started tagging specific types of errors we encountered, missing interactions, incorrect relationships, redundancy issues, etc.''} This helped improve validation approaches for recurring issues. Documentation also included records of decisions and resolutions. Participant A noted, \textit{``We keep a record of all the issues we encounter, along with the actions we take to resolve them. This is useful for audits and helps refine our internal guidelines.''} These records supported compliance and improved validation over time.

\subsubsection{Hybrid Approach}

Drawing from the \textit{Hybrid Approach} code in our codebook, organizations mitigated explainability limitations by integrating AI automation with human oversight. Participant C described efforts to improve tool performance by adjusting inputs: \textit{``We tried rewording certain requirements to see if it led to better outputs.''} However, results were inconsistent, reinforcing the need for human validation. As Participant A explained, \textit{``We no longer rely on automated outputs without verifying them first. Instead, we use a cautious, hybrid strategy, automation where it helps, but always with human oversight.''}

\subsubsection{Traceability Matrix}

Although AI tools already take requirements as input, several participants used traceability matrices not to ensure basic input-output mapping but to compensate for the tool’s lack of internal reasoning transparency. They manually linked each element in the generated diagram back to the corresponding requirement to detect missing or misrepresented content. As Participant C explained, \textit{``We cross-referenced every output from Diagramly.ai with our internal requirements traceability matrix.''} This process helped identify cases where important behaviors or constraints were omitted, even though relevant requirements had been supplied. Unlike general documentation, which logs issues over time, traceability matrices served as a validation mechanism to expose what the tool should have included but did not, offering a partial workaround for absent justifications.

While these strategies help mitigate explainability issues, they also introduce inefficiencies that disrupt workflows, extend project timelines, and create compliance risks.

\subsection{\textbf{RQ4}: Impacts on Workflow, Collaboration, and Projects}

\subsubsection{Workflow Disruption}

Drawing from the \textit{Workflow Disruption} code in our codebook, participants described workflow disruption as a consequence of non-explainable outputs, primarily due to added validation steps and resulting time delays. Participant G shared, \textit{``A planned two-week design phase took nearly six weeks because of all the manual validation required.''} These delays disrupted the development process, causing cascading setbacks and cost overruns. Unclear outputs also led to team conflicts. As Participant E noted, \textit{``When one engineer assumes the tool's output is correct, but another questions it, you end up with internal conflicts and delays.''} Resolving these issues required additional discussions and reviews, further slowing progress.

\subsubsection{Team Misalignment}

Based on the \textit{Team Misalignment} code from our codebook, with subcodes (\textit{Interpretation Variance} and \textit{Conflict Increase}), participants reported that non-explainable AI outputs caused misalignment and communication issues within teams. The lack of transparency led to frequent miscommunication. As Participant E noted, \textit{``AI tools speed up design generation, reducing back-and-forth between teams. But when outputs are unclear, they create more miscommunication than they prevent.''} Different teams often interpreted the same output differently. Participant A explained, \textit{``When an output is unclear, it creates misalignment. Designers might interpret a flagged interaction one way, while requirements engineers see it differently.''} Participant C added, \textit{``When a tool generates a UML or sequence diagram, different team members interpret it differently.''} A gap also emerged between management and engineers. Participant I remarked, \textit{``Management loved it, AI-generated UML diagrams in minutes. But management doesn’t use these diagrams. Engineers do.''} This led to misunderstandings about tool usefulness. These conflicting interpretations led to delays. Without justifications, teams held extra meetings to resolve disagreements. As Participant A recalled, \textit{``One project had multiple heated debates because the tool lacked output justifications, leaving everyone to rely on their own judgment.''} Participant C concluded, \textit{``This creates confusion, leading to longer discussions and disagreements on system function.''}

\subsubsection{Stakeholder Communication}

Participants reported that the lack of explainability complicated stakeholder interactions, as reflected in the \textit{Stakeholder Communication} code from our codebook. Management often equated faster diagram generation with efficiency, overlooking quality and validation concerns. Regulatory stakeholders, however, required clear traceability, which the tools could not provide. This gap forced teams to create supporting documentation, explain validation steps, and manage expectations. Participant F noted the difficulty of communicating tool-generated designs to external stakeholders: \textit{“We don’t even show the AI-generated outputs to external stakeholders because we can’t explain them properly.”} Without the ability to justify outputs, teams were often unable to present or defend designs in review meetings or audits, straining communication with compliance officers and project managers.

\subsubsection{Regulatory Challenges}

As captured under the \textit{Regulatory Challenges} code from our codebook with the subcode (\textit{Regulatory Challenge}), participants reported difficulties validating tool outputs that lacked clear reasoning. As Participant E noted, \textit{``In our compliance-driven projects, we have to validate every requirement against regulatory standards like ISO 13485. If an AI tool suggests a modification but doesn't explain why, we can't just accept that blindly.''} This additional effort often offset expected efficiency gains. In safety-critical fields like automotive, the stakes were higher. Participant C stated, \textit{``Safety regulations are strict. We can't afford to have incorrect diagrams leading to faulty implementations.''} Tool opacity also conflicted with audit traceability, requiring manual verification and extra documentation. Participant E added, \textit{``AI tools automate tasks without considering how humans interpret outputs. But in industry, transparency matters more than automation.''} As a result, teams in high-risk domains prioritized compliance over speed.

\subsubsection{Project Risk}

The lack of explainability introduced safety and financial risks, as captured under the \textit{Project Risk} code from our codebook. Undetected errors in tool-generated artifacts often led to production failures, especially in safety-critical domains where minor mistakes had major consequences. Tools that missed regulatory requirements or omitted key safety features posed compliance risks. As Participant B noted, \textit{``That's one of the biggest risks, when the tool's output looks correct but actually omits something important because it didn't fully understand the regulatory aspect.''} These errors increased costs, delayed projects, and required expensive rework.

\subsubsection{Safety Threat}
The lack of explainability in AI tools posed safety concerns in reliability-critical industries, as captured under the \textit{Safety Threat}.
 In medical device development, participants reported cases where tools omitted essential timeout mechanisms and alert escalation procedures, which can directly impact patient safety. As Participant E noted, \textit{``the tool had assumed a default timeout period before escalating alerts,''} a major risk since \textit{``in medical software, every second counts.''} Teams could not verify whether critical safety requirements were properly understood and applied. 

 To reduce reliance on manual validation and improve trust, practitioners identify key improvements, including contextual justifications and compliance-aware design checks.

\subsection{\textbf{RQ5}: Desired Features for Improvement}

To address explainability challenges, practitioners proposed concrete improvements mapped to specific issues. Each proposed feature targets specific challenges: contextual justifications and automated source tracing address validation and trust gaps (RQ2), compliance support mitigates regulatory risks (RQ4), while collaborative features and real-time feedback enhance team coordination and workflow reliability (RQ4). These recommendations provide a practitioner-driven roadmap for improving tool transparency, usability, and adoption in RE contexts.

\subsubsection{Contextual Justifications}

The need for AI tools to provide clear explanations for their outputs emerged as a key concern, captured under the \textit{Contextual Justifications} code with subcodes \textit{Domain Relevance} and \textit{Decision Clarity}. Current tools often generate design artifacts without disclosing their reasoning, forcing teams to reverse engineer the outputs. Rather than speeding up workflows, this lack of transparency introduces delays. Effective contextual justifications should operate at multiple levels: at the component level, explaining why elements are included or omitted; at the system level, aligning designs with key requirements; and at the validation level, clarifying decision logic to reduce manual investigation. To be useful, justifications must reflect domain-specific terminology, standards, and constraints. Generic explanations lack credibility in regulated contexts. As Participant D noted, \textit{``If a tool suggests removing a synchronization step, it should say, ‘This step may be unnecessary based on efficiency principles, but in IEC 61131-3 systems, redundancy is required for safety verification.’''} Decision clarity is also critical. When tools alter interactions or class relationships without justification, engineers cannot reliably assess correctness, increasing validation effort and reducing trust, especially in safety-critical domains.

\subsubsection{Source Tracing (Automatic)}

Participants identified automatic source tracing, linking generated outputs to specific input requirements, as the most critical feature for improving explainability. Captured under the \textit{Source Tracing (Automated)} code, this capability addresses the need to understand tool decisions without reconstructing logic manually. As Participant E noted, \textit{``If the tool modifies a design element, I need to trace that change back to the exact sentence or pattern in the input data that triggered it.''}

Several participants envisioned interactive systems where clicking on a component would reveal its source. Participant F explained, \textit{``You could click on a component and have the tool explain why it was placed there, which requirement it's linked to, and how it aligns with safety regulations.''} This would enable quick validation instead of manual tracing.

Such tracing would help validation, reduce ambiguity, support audits, and build trust. As Participant A stated, \textit{``This feature could save us days, or even weeks, of work.''} In regulated domains, this level of transparency is essential for compliance and review readiness.

\subsubsection{Validation and Compliance Support}

Practitioners emphasized the need for AI tools to support validation and compliance, especially in regulated industries. Aligned with the \textit{Validation/Compliance Support} code and \textit{Compliance Report} subcode, this reflects the demand for outputs that align with standards like ISO 13485, ISO 26262, and DO-178C. Current tools lack such checks, forcing extensive manual validation and increasing non-compliance risk. A preferred solution is to embed standard-specific checks into the design process, including real-time validation, alerts for regulatory gaps, and compliance reports linking outputs to relevant standards. These features could reduce manual effort and errors. As Participant C noted, \textit{``If the tool could flag missing compliance elements as we generate designs, we wouldn’t have to spend days manually checking everything.''} By aligning outputs with regulatory expectations, such support would enhance trust and remove a key barrier to tool adoption in compliance-driven environments.

\subsubsection{Collaborative Features}

Most tools are designed for individual users, which limits team coordination. Based on the \textit{Collaborative Features} code and its subcodes (\textit{Annotation Tools}, \textit{Version Control}, \textit{Discussion Threads}), we identify three capabilities to improve team workflows:

\textbf{Annotation Tools:} Allowing stakeholders to provide feedback directly on generated artifacts would improve collaboration. Engineers and compliance specialists could flag issues or validate design choices without modifying the original artifact, ensuring transparency.

\textbf{Version Control:} Tracking changes in requirements and designs is essential for maintaining accuracy, particularly in regulated industries. Version control would provide context for modifications, support audits, and allow teams to revert to previous versions as needed.

\textbf{Discussion Threads:} Keeping design-related discussions within the tool would reduce reliance on external platforms. Linking discussions to specific artifacts would improve decision-making and maintain  a structured design rationale.

\subsubsection{Real-Time Feedback}

Current tools operate in static mode, requiring full regeneration for any change, which slows iteration. Based on the \textit{Feedback Loop} code (\textit{User Correction}, \textit{Algorithm Refinement}), we identify the need for interactive, real-time adjustment. Real-time feedback would let engineers see the effect of changes instantly, whether updating requirements, testing design alternatives, or tuning ambiguity thresholds, enabling faster iteration and better understanding of tool logic. As Participant B noted, \textit{``If we could see real-time updates, it would eliminate so much back and forth.''} In regulated domains, immediate validation would catch compliance issues early and reduce rework. Rather than debating theoretical changes, teams could test ideas collaboratively. By shifting AI tools from static document generators to interactive assistants, real-time feedback would make them more practical for engineering workflows, supporting iterative design, compliance, and team-based decision-making.

\subsubsection{User-Centered Design}

AI tools often emphasize technical performance over usability, which limits their effectiveness in real-world engineering workflows. Adopting a user-centered design can improve adoption by aligning tools with practitioner needs. As Participant A noted, \textit{``I’d encourage researchers and developers to focus more on user-centered design.''} Providing flexible, role-specific views for architects, developers, and compliance specialists can improve communication and support seamless integration into existing processes.

\section{Discussion}

Our findings reveal systematic challenges that prevent AI tools from achieving widespread adoption in regulated industries, despite their technical capabilities.

\subsection{Balancing Explainability and Efficiency}

While AI tools can accelerate RE tasks, a lack of transparency creates unexpected inefficiencies. Practitioners reported validation tasks taking three times longer than planned due to insufficient explainability. Automation without transparency does not yield meaningful efficiency gains, as regulated industries cannot omit validation steps. Therefore, explainability must be a core feature to achieve real efficiency improvements.

\subsection{Trust as a Requirement, Not an Option}
Trust significantly affects AI tool adoption in high-risk sectors. Without transparent reasoning, practitioners are reluctant to rely on AI-generated outputs, leading to their exclusion from formal documentation and stakeholder rejection. Trust requires transparency through source traceability, contextual explanations, and compliance verification to validate outputs against industry standards.

\subsection{Challenges with Domain-Specific Knowledge}

AI tools struggle with domain-specific terminology, architectures, and regulations. Misinterpretation of key terms, misaligned outputs, and overlooked regulations result from their broad training datasets. Customized or industry-specific AI tools, trained on relevant data and adaptable to organizational needs, are necessary for effective domain-specific use.

\subsection{Human-AI Collaboration Over Full Automation}
AI tools should support, not replace, human decision-making in requirements analysis. A hybrid approach, where AI-generated artifacts are refined by human experts, balances speed and accuracy. Enhancing interaction features such as annotation tools, discussion integration, and real-time updates can improve collaboration, emphasizing the role of AI tools as supportive assistants.

\subsection{Regulatory and Compliance Considerations}
Explainability in regulated industries is mandatory, not optional. Current AI tools' lack of transparency creates additional documentation and manual validation efforts. Tools should automate compliance checks, flag potential issues, and produce traceable documentation alongside outputs to streamline regulatory validation.

\subsection{Economic and Organizational Considerations}

Adopting AI tools involves hidden financial and organizational challenges, including extensive validation and documentation requirements. Organizations should consider alignment with workflows, compliance, and collaboration practices beyond technical capabilities. Adequate planning for process adjustments and user training ensures that AI tools provide genuine benefits.

\subsection{Recommendations for Future AI Tools in RE}

Our findings suggest that future AI tools for RE should prioritize explainability, domain adaptation, collaboration, and regulatory compliance to improve usability and trust, especially in regulated industries.

\subsubsection{Improve Explainability and Source Tracing} 

Current AI tools lack transparency, making validation difficult. Future tools should provide \textit{contextual justifications} for design choices and \textit{source tracing} to link outputs to original requirements, reducing manual validation effort.

\subsubsection{Support Domain-Specific Adaptation and Compliance Checks} 

Tools often fail to handle industry-specific terminology and regulations. Future tools should support \textit{customizable domain ontologies} and \textit{built-in compliance checks} to align outputs with engineering needs and reduce manual validation.

\subsubsection{Enable Real-Time Feedback and Iterative Refinement} Future AI tools should provide \textit{real-time feedback} and make updating design artifacts instant when requirements change. This would help teams refine outputs efficiently without regenerating entire models.

\subsubsection{Improve Collaboration Features} 

Future AI tools should support \textit{annotation tools, version control, and discussion threads} to facilitate team-based workflows. These features would help stakeholders track changes, document reasoning, and resolve issues within the tool, reducing reliance on external communication platforms.

\subsubsection{Integrate with Existing Workflows} 
AI tools should fit into engineering workflows without the need for major process changes. \textit{Flexible viewing options} should tailor information to user roles, while \textit{customizable detail levels} should let users control displayed information at different review stages. 

These improvements would help RE teams produce reliable, traceable, and compliant design artifacts, making AI tools more practical for regulated and safety-critical industries.

\section{Threats to Validity}

We evaluate the limitations of our study using Guba and Lincoln’s \cite{lincoln1985naturalistic} trustworthiness framework.

\subsection{Credibility}
Our study relies on semi-structured interviews, so there is some risk of recall bias. To mitigate this, we asked for specific examples and used investigator triangulation, with two independent coders analyzing the data (Cohen’s Kappa = 0.76). As we did not observe participants in situ, our findings are based on self-reported data, which may limit validation. We did not conduct a formal pilot test or external expert validation of the interview guide, which may affect construct validity. To mitigate this, we iteratively refined questions during early interviews and sought in-depth responses to clarify ambiguities.

\subsection{Transferability}
Participants were drawn from regulated industries with strict explainability demands, which may not generalize to less-regulated domains. To aid contextualization, we report participant roles and domain expertise, though our sample may not cover all regulatory or geographic variations.

\subsection{Dependability}

We followed a standardized interview protocol and a systematic coding process, documented in a public codebook. We reached data saturation after six interviews, which indicates stability. As technologies evolve, participant experiences may reflect tool generations that differ from future implementations. We maintained an audit trail and used two independent coders to reduce bias. While selection bias is possible, we included participants with diverse experience levels, including those who had discontinued tool use. Nonetheless, some self-selection bias may remain.

\section{Conclusion}
In this paper, we examined the explainability challenges of AI tools in RE, highlighting their impact on adoption, validation efforts, and regulatory compliance in safety-critical industries. Through interviews with practitioners, we identified key barriers, including the need for extensive manual validation, trust concerns, domain-specific limitations, and workflow disruptions. Participants emphasized that current AI tools require additional manual validation efforts, which often negate the intended automation benefits. They also highlighted that a lack of explainability directly reduces stakeholder trust and complicates compliance. To address these issues, we outlined essential improvements, such as automated source tracing, clear contextual justifications, built-in compliance validation, and collaborative features to improve team alignment and stakeholder communication. Our findings emphasize that explainability is essential, not optional, for integrating AI tools into RE workflows, particularly in regulated environments. Future research should empirically evaluate these proposed improvements, explore domain-specific adaptations, and assess the long-term impact of enhanced explainability on RE processes and overall project success.

\bibliographystyle{IEEEtran}
\bibliography{simple}
\end{document}